\documentclass[aps,prb,twocolumn,amsmath,amssymb]{revtex4}
\usepackage{epsfig}
\usepackage{graphicx}
\usepackage{multirow}
\usepackage{tabularx}

\newcommand{\be}{\begin{equation}}
\newcommand{\ee}{\end{equation}}
\newcommand{\mean}[1]{\langle #1 \rangle}

\begin{document}

\title{Charge and spin transport through a ferromagnet/insulator/unconventional superconductor junction}
\author{Gaetano Annunziata, Mario Cuoco, Paola Gentile,  Alfonso Romano, Canio Noce}
\affiliation{CNR-SPIN, I-84084 Fisciano (Salerno), Italy \\
Dipartimento di Fisica ``E. R. Caianiello'' Universit\`a di
Salerno, I-84084 Fisciano (Salerno), Italy}

\begin{abstract}
We analyze the charge and spin transport through a ballistic ferromagnet/insulator/superconductor junction by means of the Bogoliubov�-de Gennes equations. For the ferromagnetic side we assume that ferromagnetism may be driven by an unequal mass renormalization of oppositely polarized carriers, i.e. a spin
bandwidth asymmetry, and/or by a rigid splitting of up-and down-spin electron bands, as in a standard Stoner ferromagnet, whereas the superconducting side is assumed to exhibit a $d$-wave symmetry of the order parameter, which can be pure or accompanied by a minority component breaking time-reversal symmetry.
Several remarkable features in the charge conductance arise in this kind of junction, providing useful information about the mechanism of ferromagnetism in the ferromagnetic electrode, as well as of the order parameter symmetry in the superconducting one. In particular, we show that when a time-reversal symmetry
breaking superconductor is considered, the use of the two kinds of ferromagnet mentioned above represents a valuable tool to discriminate between the different superconducting mixed states. We also explain how this junction may mimic a switch able to turn on and off a spin current, leaving the charge conductance unchanged, and we show that for a wide range of insulating barrier strengths, a spin bandwidth asymmetry ferromagnet may support a spin current larger than a standard Stoner one.
\end{abstract}

\date{\today}

\maketitle

\section{INTRODUCTION}
Transport in itinerant ferromagnet/insulator/super\-conductor (F/I/S) junctions is a fundamental issue in condensed matter physics for its deep implications in electronics and spintronics,~\cite{spintronicsreview} and for the opportunity it offers to test the physical properties of ferromagnetic and superconducting materials via point contact~\cite{Soulen98,Mazin99} or scanning tunneling~\cite{Tedrow94,Bode03} measurements. Moreover, this type of hybrid structure may serve as a playground for a wealth of interesting quantum mechanical effects pertaining to the interplay between spin and charge degrees of freedom. In fact, beyond the possibility of a direct estimation of the gap magnitude in conventional superconductors, tunneling conductance measurements offer the opportunity to probe also the superconducting order parameter symmetry in unconventional superconductors. This property has made this kind of measurements fundamental in finding clues about the symmetry of the new families of superconductors recently discovered, for which there is a general consensus that they cannot be considered as conventional. For example, one of the strongest evidences supporting $d$-wave symmetry for high-$T_c$ cuprates is the zero bias conductance peak (ZBCP) revealed in $ab$-plane tunneling conductance from normal metals.~\cite{DeutRev} In some cuprates, such as for instance YBa$_2$Cu$_3$O$_{7-\delta}$,~\cite{Covington97,YBCO} the existence of a subdominant component in the order parameter possibly breaking time-reversal symmetry is still a matter of debate and, in this respect, exploiting the interplay between magnetism and superconductivity in tunneling experiments is one of the standard routes to investigate this issue. Generally, by using a ferromagnetic electrode in tunneling experiments it is possible to change the relative contributions of up and down electrons to the total density of states or, in the half-metal limit, to isolate a single spin channel.

To interpret the large amount of tunneling experiments performed on F/I/S junctions involving an unconventional superconductor,~\cite{TunnHF,TunnOrganics,TunnSrRuO,TunnFullerene,TunnBC,TunnBi,TunnFS,TunnFeAs} fundamental theories of transport, such as in particular the one by Blonder, Tinkham and Klapwijk (BTK),~\cite{BTK} have been suitably extended to take into account all possible symmetries of the superconducting order parameter. In this context, the ferromagnetic electrode has been  predominantly described within the Stoner model, relying on the assumption that the bands associated with the two possible electron spin orientations have identical dispersion, but are rigidly shifted in energy by the exchange interaction. However, Stoner model may prove to be insufficient to describe real ferromagnets because many terms deriving from Coulomb repulsion are eliminated from the full Hamiltonian, although in some situations their contribution can be important.~\cite{Abinitio,Fazekas}

The complexity of ferromagnetism in metals is testified by the wide range of manifestations it exhibits in nature. As relevant examples of this variety, we mention the ferromagnetic transition metals Fe, Co, and Ni and their alloys,~\cite{Wohlfarth80} weak metallic ferromagnets such as ZrZn$_2$~\cite{ZrZnJap,ZrZnNature} and Sc$_3$In,~\cite{Matthias61,HirschScIn} colossal magnetoresistance manganites such as La$_{1-x}$Sr$_x$MnO$_3$,~\cite{Schiffer95} and rare earth hexaborides such as EuB$_6$.~\cite{Matthias68,HirschEuB} Therefore, when theoretically modelling F/I/S junctions, it may be important to assume for the magnetism in the F electrode microscopic scenarios other than the Stoner one. Among them, of peculiar interest is a form of itinerant ferromagnetism driven by a gain in kinetic energy deriving from a spin dependent bandwidth renormalization, or, equivalently, by an effective mass splitting between up- and down-spin carriers.~\cite{Zener51,Hirsch,Campbell88,Okimoto95,Higashiguchi05,McCollam05} The interplay of superconductivity with this kinetically driven ferromagnetism has been recently shown to originate different features compared to the Stoner case, concerning the phenomena of coexistence, proximity and transport. More precisely, we have studied the occurrence of the coexistence of ferromagnetism and $s$-wave singlet superconductivity within a model where the magnetic moments are due to a kinetic exchange mechanism, and we have shown that the depaired electrons play a crucial role in the energy balance, and that when their dynamical effect is such that to undress the effective mass of the carriers which participate in the pairing, a coexisting ferromagnet-superconducting phase can be stabilized.~\cite{Cuoco03} Then, we have exactly solved an extended version of the reduced BCS model for particles that get paired in the presence of a polarization arising from spin dependent bandwidths and we have calculated the ground-state phase diagram in the full parameter space of the pair coupling and the bandwidth asymmetry as a function of filling for different types of spectrum topologies.~\cite{Ying} We have also investigated the
proximity effect within a junction made of an unconventional superconductor and a ferromagnet in the clean limit with high barrier transparency, and we have shown that the two above-mentioned mechanisms for ferromagnetism lead to different features as concerns the formation at the interface of dominant and sub-dominant superconducting components as well as their propagation in the ferromagnetic side.~\cite{Cuoco08} Finally, a F/I/S ballistic junction with a conventional $s$-wave superconductor has been used to distinguish whether itinerant ferromagnetism in the F electrode is due to exchange splitting or to spin-dependent mass renormalization of up- and down-spin electrons.~\cite{Annunziata} We have also shown that under appropriate conditions the spin dependent conductance of minority carriers can be larger than for majority carriers below the energy gap $\Delta_0$, and lower above it, suggesting that the junction, in a suitable range of microscopical parameters, may work as a spin-filtering device.~\cite{Sust}

In this work we carry out the investigation of the interplay between different types of ferromagnetism and superconductivity analyzing charge and spin transport through a ballistic F/I/S junction where various unconventional symmetries for the superconducting electrode are considered. The problem is handled by solving the Bogoliubov�-de Gennes (BdG) equations~\cite{BdG} within an extended Blonder-Tinkham-Klapwijk approach, here formulated for a two-dimensional F/I/S junction. As it is well known, this method has been generalized in the last years to take into account higher dimensionalities, anisotropic forms of the superconducting order parameter, different Fermi energies for the two sides of the junction, and a spin--flip interfacial scattering.~\cite{Kashiwaya99,ZuticValls,Dong,Zhu00,Zhu99,Barsic,Stefan1,Linder,deJong,Tanaka_transp_d,Tanaka_transp_p,YokoyamaNCSC,mariospinactive} We investigate the behavior of the charge and the spin conductance, revealing several noteworthy features arising from the interplay between unconventional superconductivity and each of the two kinds of ferromagnetism specified above. The differences emerging in the two cases are shown to provide relevant indications on the physical properties of the materials constituting both the ferromagnetic and the superconducting electrode of the junction. Moreover, we also show that the behavior of the charge conductance in the case of pure $d$-wave materials is different from that found when a minority component breaking time-reversal symmetry (BTRS) is also present. In this case, the new features emerging around zero-bias voltage exhibit a different behavior depending on whether a Stoner or a mass mismatch ferromagnet is considered, thus providing a clear indication on the nature of the microscopic mechanism underlying ferromagnetism in the F layer. We would like to notice that our assumption of a bulk character of these BTRS components is justified by the two-dimensional character of the junction that we analyze. Indeed, with this two-dimensional geometry, a nodeless broken time reversal symmetry state may appear throughout the S side of  the junction, and this is consistent with the BTRS $d_{x^2-y^2}+is$ or $d_{x^2-y^2}+id_{xy}$ combinations here assumed.

We also report on the effect of the mass asymmetry on spin conductance for conventional and unconventional superconducting electrodes and we show that under specific conditions the mass mismatch greatly enhances spin current, so that a spin bandwidth asymmetry ferromagnet can lead to a spin current much larger than the that produced by a Stoner ferromagnet, at the same polarization. Hence, we explain how a F/I/S junction can work as a switch able to turn on and off a spin current, leaving the charge current unchanged.

The paper is organized as follows. In Section II we formulate the microscopic model and the related method of solution. In Section III we present the results, discussing them in three different Subsections concerning: (A) the magnetization in the ferromagnetic side, (B) the charge conductance through a junction with a superconductor having a pure $d_{x^2-y^2}$-wave symmetry or a broken time-reversal symmetry of $d_{x^2-y^2}+is$ or $d_{x^2-y^2}+id_{xy}$ type, and (C) the spin conductance through a junction with conventional and unconventional superconductors. Finally, Section IV is devoted to the conclusions.

\section{MODEL AND FORMALISM}

\begin{figure}
\begin{center}
\includegraphics[width=0.4\textwidth]{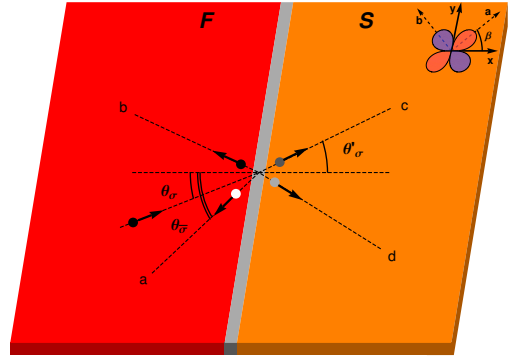}
\caption{(Color online) Scheme of the planar F/I/S junction
analyzed in the paper. Here, $\theta_\sigma$, $\theta_{\bar{\sigma}}$,
and $\theta_\sigma '$ are injection, Andreev reflection, and
transmission angles, respectively,  for electrons and quasiparticles with spin
$\sigma$. $\beta$ is the angle formed by the crystallographic $a$
axis of a $d$-wave superconductor with the $x$ axis.}
\label{sketch}
\end{center}
\end{figure}

The system under study is built up of two semi-infinite layers connected by an infinitely thin insulating barrier resulting in an interfacial scattering potential of the form $V(\mathbf{r})=H\delta(x)$. We choose an interface lying along the $y$ direction at $x=0$ (see Fig.~\ref{sketch}) so that the region $x<0$ (from now on the F side) is occupied by an itinerant ferromagnet (a Stoner or a spin bandwidth asymmetry ferromagnet, or a combination of the two), while the region $x>0$ (from now on the S side) is occupied by a singlet superconductor (so there is no need to specify the spin quantization axis). 
We point out that though in the following we refer to free particle-like spectra of parabolic type for which the concept of bandwidth is in principle ill-defined, we nonetheless imagine to link this description to some effective one-band tight-binding model, allowing to relate the inverse of the mass of the carriers to the width of the effective bands where itinerancy takes place. In this way, a bandwidth asymmetry is generated  by simply assuming different values of the masses for up- and
down-spin electrons.

We describe the excitations propagating through the junction by means of the single-particle Hamiltonian
\begin{eqnarray}
H_0^\sigma & = & \left[
-\hbar^2\mathbf{\nabla}^2/2m_\sigma-\rho_\sigma U
-E_F\right]\Theta(-x) \nonumber \\ && +
 \left[ -\hbar^2 \mathbf{\nabla}^2/2m'-E_F' \right]\Theta(x)+V(\mathbf{r})\; ,
\label{spHam}
\end{eqnarray}
where $\sigma=\uparrow,\downarrow$, $m_\sigma$ is the effective mass for $\sigma$-polarized electrons in the F side, $\rho_{\uparrow(\downarrow)}=+1(-1)$, $U$ is the exchange interaction, $E_F$ is the Fermi energy of the ferromagnet,
$\Theta(x)$ is the unit step function, $m'$ and $E_F'$ are the quasiparticles effective mass and the Fermi energy for the
superconductor, respectively.

The BdG equations read as
\be
\left( \begin{matrix} H_0^\sigma && \Delta \\
\Delta^\ast && -H_0^{\bar{\sigma}} \end{matrix} \right) \left(
\begin{matrix} u_\sigma \\ v_{\bar{\sigma}} \end{matrix} \right)
=\varepsilon \left( \begin{matrix} u_\sigma \\ v_{\bar{\sigma}}
\end{matrix} \right), \,\sigma=\uparrow,\downarrow\quad,
\label{eigensystem}
\ee
where $\bar{\sigma}=-\sigma$ and $\left(u_\sigma,v_{\bar{\sigma}}\right)\equiv\Psi_\sigma$ is the energy eigenstate in the electron-hole space associated with the eigenvalue $\varepsilon$ (excitation energies are measured from the Fermi level). Eqs.~(\ref{eigensystem}) admit an analytical solution in the approximation of a rigid superconducting pair potential, i.e. $\Delta(\mathbf{r})=\Delta(\theta')\Theta(x)$, where $\theta'$ is the angular variable for the S side (see Fig.~\ref{sketch}). The Hamiltonian invariance under $y$-directed translations permits to factorize the part of the eigenstate parallel to the interface, i.e. $\Psi_\sigma(\mathbf{r})= e^{i\mathbf{k}_\parallel\cdot\mathbf{r}}\psi_\sigma(x)$, reducing the effective dimensionality of the problem.

Looking at Fig.~\ref{sketch}, we observe that at the interface four scattering processes are possible for an electron injected from the F side with spin $\sigma$ and momentum
$\mathbf{k}_\sigma^+$ ($k_\sigma^+=\left[\left(2m_\sigma/\hbar^2\right)\left(E_F+\rho_\sigma U+\varepsilon\right)\right]^{1/2}$): $a)$ Andreev reflection (AR) resulting in a hole with momentum $\mathbf{k}_{\bar{\sigma}}^-$ ($k_{\bar{\sigma}}^-= \left[\left(2m_{\bar{\sigma}}/\hbar^2\right)\left(E_F+\rho_{\bar{\sigma}} U-\varepsilon\right)\right]^{1/2}$) belonging to the opposite spin band and a Cooper pair transmitted in the superconductor; $b)$ normal reflection; $c)$ transmission as electron--like quasiparticle with momentum $\mathbf{k}_\sigma '^+$ ($k_\sigma '^+= \left[\left(2m'/\hbar^2\right)\left(E_F'+\sqrt{\varepsilon^2-|\Delta_{\sigma +}|^2}\right)\right]^{1/2}$); $d)$ transmission as hole--like quasiparticle with momentum $\mathbf{k}_\sigma '^-$ ($k_\sigma '^-=
\left[\left(2m'/\hbar^2\right)\left(E_F'-\sqrt{\varepsilon^2-|\Delta_{\sigma -}|^2}\right)\right]^{1/2}$),
where $\Delta_{\sigma \pm}=|\Delta_{\sigma \pm}| \ e^{i \phi^{\pm}_{\sigma}}$ is the pair potential felt by electron-like ($+$) and hole-like ($-$) quasiparticles. We notice that the spin dependence of $\Delta_{\sigma \pm}$ comes out from the different trajectories followed by up- and down-spin quasiparticles. Which of these processes actually takes place depends on the energy, momentum and spin orientation of the incoming electrons, as well as on the interfacial barrier strength, the polarization in the F side and the symmetry of the superconducting order parameter in the S side.

For standard low-biased F/I/S junctions, one has $E_F,E_F'\gg(\varepsilon,|\Delta|)$, so that one can apply the Andreev approximation~\cite{Andreev64} and fix the momenta on the Fermi surfaces. In this case the solutions of BdG equations for the two sides of the junction can be written as
\begin{eqnarray}
\psi_\sigma^F(x)=e^{ik_{\sigma,x}^Fx}\left(\begin{matrix} 1\\0
\end{matrix} \right)+ a_\sigma
e^{ik_{{\bar{\sigma},x}}^Fx}\left(\begin{matrix} 0\\1 \end{matrix}
\right)+
b_\sigma e^{-ik_{\sigma,x}^Fx}\left(\begin{matrix} 1\\0 \end{matrix} \right)\\
\psi_\sigma^S(x)=c_\sigma e^{ik_{\sigma,x}'^Fx}\left(
\begin{matrix} u_+\\
e^{-i\phi_+}v_+
\end{matrix} \right)+ d_\sigma
e^{-ik_{\sigma,x}'^Fx}\left(\begin{matrix}
e^{i\phi_-}u_-\\
v_- \end{matrix} \right)
\end{eqnarray}
where
$$
u_\pm  = \sqrt{\frac{\varepsilon\pm\sqrt{\varepsilon^2-|\Delta_{\sigma \pm}|^2}}{2\varepsilon}}
$$
$$
v_\pm  =
\sqrt{\frac{\varepsilon\mp\sqrt{\varepsilon^2-|\Delta_{\sigma \pm}|^2}}{2\varepsilon}}
\; ,
$$
and the superscript F in the wave-vectors denotes that they are taken on the Fermi surfaces.

The boundary conditions at the interface allow for the calculation of the probability amplitude coefficients $a_\sigma$, $b_\sigma$, $c_\sigma$, $d_\sigma$ for the four scattering processes. We have

\begin{subequations}\label{raccordo}
\begin{eqnarray}
\psi_\sigma^F(0)&=&\psi_\sigma^S(0) \\
\left.\frac{m_\sigma}{m'}\frac{d
u_\sigma^S}{d x}\right|_{x=0}- \left.
\frac{d u_\sigma^F}{d x}\right|_{x=0}
&=&\frac{2 H\ m_\sigma}{\hbar^2}u_\sigma^S(0)\\
\left. \frac{m_{\bar{\sigma}}}{m'}\frac{d v_{\bar{\sigma}}^S}{d
x}\right|_{x=0}-\left.
\frac{d v_{\bar{\sigma}}^F}{d x}\right|_{x=0}
&=&\frac{2 H\
m_{\bar{\sigma}}}{\hbar^2}v_{\bar{\sigma}}^S(0) \; .
\end{eqnarray}
\end{subequations}

\noindent Eq.~(\ref{raccordo}) show that the mass asymmetry explicitly renormalizes the interface
barrier strength $H$, giving rise to a dependence of this quantity on the spin of the carriers. This effect, which under
suitable conditions leads to a different behavior of these carriers across the barrier, allows to infer that the presence of spin dependent electron masses in Eq.~(\ref{raccordo}) may mimic a spin active barrier, in the sense that electrons with opposite spin feel different values of the barrier height. A junction with a mass mismatch ferromagnet can thus induce an effective spin-active interfacial effect, which for specific choices of $H$ has been shown to produce a minority--spin charge conductance component higher than the corresponding majority-spin one.~\cite{Sust} The dimensionless parameter $Z=2m'H \pi^2 /(\hbar^2 k'_F)$ everywhere in following will conveniently characterize the strength of the interfacial scattering.

The charge and spin differential conductances at $T=0$ and energy $\varepsilon$, i.e. at bias voltage $V=\varepsilon/e$, $e$ being the electron charge, are calculated from the ratio between the charge and spin fluxes across the junction and the incident fluxes at that bias. They can be easily obtained from the probabilities associated with the four processes listed above,~\cite{ZuticValls} and for each spin channel they can be written as
\begin{eqnarray}
G_\sigma(\varepsilon,\theta)= P_\sigma \left(
1+\frac{k_{\bar{\sigma},x}^F}{k_{\sigma,x}^F} \
|a_\sigma(\varepsilon,\theta)|^2-|b_\sigma(\varepsilon,\theta)|^2
\right) \\  \Sigma_\sigma(\varepsilon,\theta)= P_\sigma \left(
1-\frac{k_{\bar{\sigma},x}^F}{k_{\sigma,x}^F} \
|a_\sigma(\varepsilon,\theta)|^2-|b_\sigma(\varepsilon,\theta)|^2
\right),
\end{eqnarray}
where $\theta$ is the angle formed by the momentum of the electrons propagating from the F side with respect to the normal to the interface (see Fig.~\ref{sketch}), and the polarization $P_\sigma=n_\sigma/(n_\uparrow+n_\downarrow)$ is the fraction of
electrons occupying the $\sigma$-spin band of the metallic ferromagnet.

The measured conductances take contributions from a range of angles determined by the experimental conditions. This range is
limited from above due to the conservation of the momentum parallel component
\be
\label{snell} k_\sigma^F \sin \theta=k_{\bar{\sigma}}^F \sin \theta_{\bar{\sigma}}=k'^F \sin \theta_\sigma ' \; ,
\ee
where $\theta_{\bar{\sigma}}$ and $\theta_\sigma '$ are AR and the transmission angles, respectively, for electrons and quasiparticles with spin $\sigma$.
From this equation it is easy to verify the existence of critical angles above which these processes are no more possible, resulting in virtual AR~\cite{ZuticValls} and normal reflection, respectively. The angularly averaged differential conductances for given spin orientation are then defined as~\cite{ZuticValls}
\begin{eqnarray}
\mean{G_\sigma(\varepsilon)}=\int_{-\theta_{C}^\sigma}^{\theta_{C}^\sigma}
d\theta \ \cos \theta \ G_\sigma(\varepsilon,\theta) /
\int_{-\theta_{C}^\sigma}^{\theta_{C}^\sigma} d\theta \ \cos
\theta \\
\mean{\Sigma_\sigma(\varepsilon)}=\int_{-\theta_{C}^\sigma}^{\theta_{C}^\sigma}
d\theta \ \cos \theta \ \Sigma_\sigma(\varepsilon,\theta) /
\int_{-\theta_{C}^\sigma}^{\theta_{C}^\sigma} d\theta \ \cos
\theta ,
\end{eqnarray}
where $\theta_{C}^\sigma$ is the critical angle for the transmission of {$\sigma$-spin} electrons.

Finally, the net averaged charge and spin conductances are respectively defined as
\begin{eqnarray}
\mean{G(\varepsilon)} & = & \mean{G_\uparrow(\varepsilon)}+\mean{G_\downarrow(\varepsilon)} \label{charge cond}\\
\mean{\Sigma(\varepsilon)} & = &
\mean{\Sigma_\uparrow(\varepsilon)}-\mean{\Sigma_\downarrow(\varepsilon)}
\;  . \label{spin cond}
\end{eqnarray}

\section{RESULTS}

The results here obtained for the F/I/S junction are grouped in three following distinct Subsections concerning (A) the
magnetization in the F side, (B) the charge conductance for a $d_{x^2-y^2}$-wave or broken time-reversal states (BTRS) associated with $d_{x^2-y^2}+is$ and $d_{x^2-y^2}+id_{xy}$ pairing symmetry superconducting electrode, and (C) the spin conductance for the above choices of the order parameter and, for a comparison, for a conventional $s$-wave superconducting S-side. To appreciate the effect of mass asymmetry, we neglect Fermi energies mismatch effects and fix $E_F=E_F'$.

\begin{figure}[!th]
\begin{center}
\includegraphics[width=0.5\textwidth]{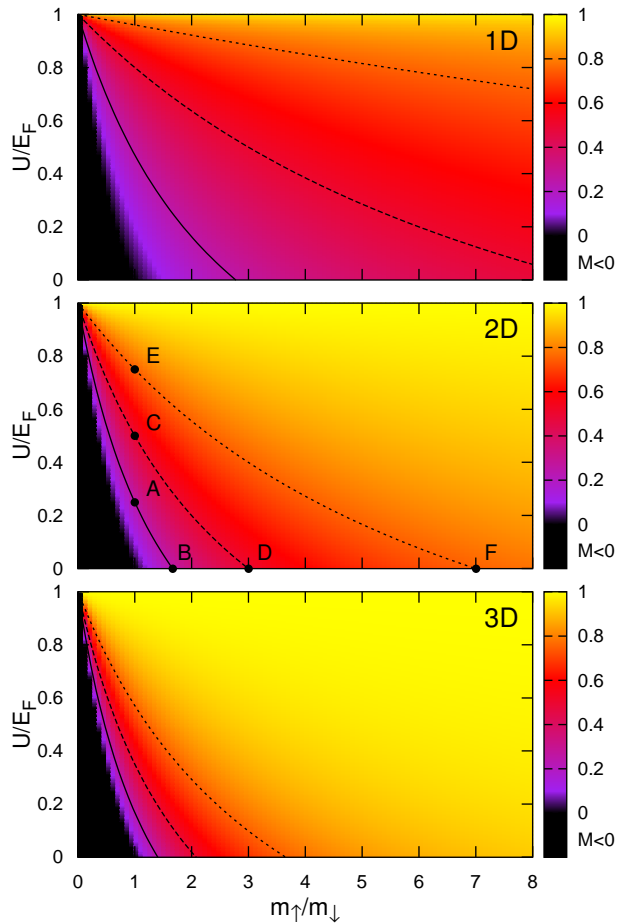}
\caption{(Color online).
Density plot of the ground state magnetization as a function of the
mass mismatch and the normalized exchange interaction, for one-,
two- and three-dimensional ferromagnetic electrodes. As shown in
the legend on the right, lighter color regions are associated with
higher values of the magnetization. For clarity, only three
iso-magnetization curves are plotted in all panels, corresponding
to $M=0.25$ (solid line), $M=0.50$ (dashed line), and $M=0.75$
(dotted line). In the middle panel, referring to the
dimensionality considered in this paper, we depict six representative points:
A and B correspond to two different microscopic states with the same macroscopic magnetization M = 0.25, A reprsenting a standard Stoner ferromagnet ($m_\uparrow/m_\downarrow=1$),
and B a purely spin bandwidth asymmetry ferromagnet
($U/E_F=0$). The same holds for the (C,D) and (E,F) couples of
points, referring to higher values of the magnetization ($M=0.50$
and $M=0.75$, respectively). The values assumed by the microscopic parameters
in the above mentioned six states are summarized in Table~\ref{table}.}
\label{parameterspace}
\end{center}
\end{figure}

\subsection{MAGNETIZATION}
The spin bandwidth asymmetry in the F side directly affects the density of states per spin orientation, and consequently the net polarization. Given the single-particle Hamiltonian (\ref{spHam}), we find that in the two-dimensional case the ground-state magnetization $M \equiv P_\uparrow-P_\downarrow$ is  given by
\be
\label{mag} M_{2D}=\frac{(X+1) Y}{X (Y-1)+Y+1}-\frac{1-X}{X
(Y-1)+Y+1}\, ,
\ee
where $X=U/E_F$ and $Y=m_\uparrow/m_\downarrow$. Eq.~(\ref{mag}) correctly reduces to known results for a pure Stoner ferromagnet when $Y\rightarrow1$.~\cite{Dong} On the other hand, when $Y\rightarrow0 (\infty)$ we precisely reproduce the half-metal limit $M\rightarrow-1 (1)$. For a fixed value of the exchange splitting, the mass mismatch enhances the net polarization for $m_\uparrow > m_\downarrow$ ($Y>1$) and hinders it the other way around ($Y<1$). The situation is illustrated in Fig.~\ref{parameterspace}, where the density plot of the magnetization at $T=0$ in the ($m_\uparrow/m_\downarrow$, $U/E_F$) parameter space is shown for one-, two- and three-dimensional ferromagnets, together with three isomagnetization curves plotted to clarify the magnetization trend. Each point corresponds to a different realization of the ferromagnetic order in the sense that the relative weights of the exchange splitting and the mass mismatch are determined by the coordinates of that point, while the value of $M$ is fixed along the isomagnetization curves (the solid, dashed and dotted lines of Fig.~2). We see that though the qualitative behavior is independent on the dimensionality, for the chosen band dispersion in the ferromagnet, one always finds $M_{3D}>M_{2D}>M_{1D}$ when evaluated for the same $m_\uparrow/m_\downarrow$ and $U/E_F$ values.

\begin{center}
\begin{table}
\begin{tabular}{p{0.1cm}p{3.4cm}cccccp{0.2cm}}
\hline\hline
& &&&&&&\\
& & $U/E_F$ &$ \quad   $& $m_\uparrow /m_\downarrow$ &$ \quad  $&$M$&\\
& &&&&&&\\
\hline
& &&&&&&\\
&A &0.25 && 1&&\multirow{2}{*}{0.25}&\\
&B &0 && 5/3 &&&\\[2ex]
&C &0.50 &&1 &&\multirow{2}{*}{0.50}&\\
&D &0 & &3&&&\\[2ex]
&E &0.75 &&1&& \multirow{2}{*}{0.75}&\\
&F &0 && 7&&&\\
\hline\hline
\end{tabular}
\caption{Values of the normalized exchange interaction $U/E_F$, the mass mismatch $m_\uparrow /m_\downarrow$ and the magnetization $M$ for the six illustrative points displayed in the middle panel of Fig.~\ref{parameterspace}.} \label{table}
\end{table}
\end{center}

\subsection{CHARGE TRANSPORT}
We have analyzed F/I/S conductance spectra in two dimensions in the entire parameter space, excluding the regions corresponding to $M<0$ (indicated in black in Fig.~\ref{parameterspace}) since they are mirror images of those with positive $M$, assuming that $m_\uparrow/m'=m'/m_\downarrow$ for $Y>1$. We notice that with this choice critical angles for AR and transmission exist only for majority electrons.
In the following Subsections we discuss the results for the six representative points highlighted in Fig.~\ref{parameterspace}, which correspond to a pure Stoner ferromagnet (STF), i.e. $m_\uparrow/m_\downarrow=1$, and a pure spin bandwidth asymmetry ferromagnet (SBAF), i.e. $U/E_F$=0, for three different values of the magnetization $M=0.25, 0.50, 0.75$ (the corresponding values of $m_\uparrow/m_\downarrow$ and $U/E_F$ are reported in Table~\ref{table}). F/I/S conductance spectra will be shown for various symmetries of the superconducting order parameter, emphasizing the differences in transport between STF/I/S and SBAF/I/S junctions. We finally notice that spectra change continuously as one moves along an isomagnetization curve from a
point corresponding to a STF to a point corresponding to a SBAF.

\subsubsection{F/I/$d_{x^2-y^2}$ JUNCTION}
It is well known~\cite{HuZBCP,KashiwayaZBCP} that in N/I/S junctions involving a normal metal and a $d_{x^2-y^2}$
superconductor, a zero-bias conductance peak (ZBCP) develops in the tunneling limit, this peak becoming narrower and narrower as increasing values of the interfacial barrier strength are considered. This ZBCP is the consequence of the presence of an
Andreev bound state~\cite{ABS} (ABS) at the Fermi energy, induced by the change in sign of the pair potential across line nodes. It implies that electron-like and hole-like quasiparticles specularly reflected at the interface always find the ``right" sign of the pair potential to be Andreev reflected. In this case the ABS is at the same energy for every quasiparticle trajectory, i.e. for every angle $\theta$. When the normal metal in the junction is replaced by a STF, the ZBCP is lowered because of the presence of the ferromagnetic polarization which inhibits ARs and can be splitted in two sub-peaks developing symmetrically at finite energies,~\cite{Dong,Zhu00} depending on interfacial scattering strength.
The splitting of the ZBCP is clearly visible in the angle-resolved charge conductance, while in the angle-averaged one it is distinguishable only for high magnetization. However, when the interface barrier strength $Z$ is reduced, this structure becomes better defined since the two peaks get more separated, though less pronounced.

\begin{figure}
\begin{center}
\includegraphics[width=0.45\textwidth]{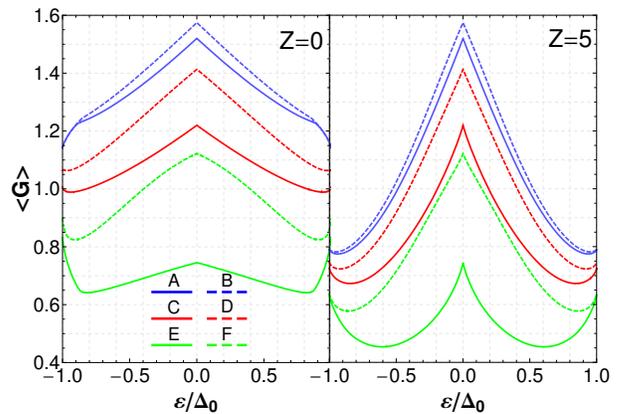}
\caption{(Color online). Averaged differential
conductance spectra for a junction with a $d_{x^2-y^2}$-wave
superconducting electrode, evaluated in the states corresponding to the six points indicated
in Fig.~\ref{parameterspace} and listed in Table~\ref{table}, in
the metallic limit $Z=0$ (left panel) and for intermediate barrier
transparency $Z=5$ (right panel).} \label{dwave}
\end{center}
\end{figure}

Now, let us investigate how this picture is modified when a SBAF is taken into account. We remind that, when the superconducting electrode has $d$-wave symmetry, the pair potential felt by electrons (+) and holes (--) is $\Delta_{\sigma,\pm}=\Delta_0 \cos[2(\theta_\sigma'\mp\beta)]$, where $\beta$ is the angle formed by the crystallographic $a$ axis of the superconductor with the $x$ axis (see Fig.~\ref{sketch}). We here fix $\beta=\pi/4$ to analyze a $d_{x^2-y^2}$-wave superconductor with line nodes perpendicular to the interface. In Fig.~\ref{dwave} we show the averaged differential conductance spectra evaluated at the six points highlighted in Fig.~2 and listed in Table I, in the limit of full transparency of the barrier (left panel) and for an intermediate value of $Z$ (right panel). Comparing the behavior of a SBAF/I/$d_{x^2-y^2}$ and a STF/I/d$_{x^2-y^2}$ junction, we find qualitative deviations in the charge conductance which become more and more significant as increasing values of the magnetization and of the barrier strength are considered (see Fig.~\ref{dwave}). It is found that with the increase of the magnetization the ZBCP is lowered rapidly and eventually smeared out in the STF case, whereas in the SBAF case the ZBCP is more robust against the polarization of the F-side.
The drop of the zero bias conductance may be attributed to the fact that for a given injection angle, when $M$ increases above a threshold, the AR processes for the incident electron with spin up is suppressed and only the AR of spin down electrons contributes to the ZBCP. We point out that this behavior can be rigorously proved considering that the ABS amplitude decreases with increasing exchange field due to the sensitivity of Andreev reflections to spin polarization, represented in BTK-type models by a suppression in the Andreev term coefficient.
This picture is slightly modified when a SBAF is considered, since in this case the barrier, according to Eq.~(6), may be spin selective, assisting the conductance of the two spin channels in a different way. This effect results into a charge conductance always larger than the one obtained in the corresponding STF case with the same magnetization $M$. Finally, we notice that with increasing $Z$, i. e. when we move from the metallic limit towards the tunneling one, the averaged charge conductance here obtained reproduces the well-known behavior previously reported in the literature.~\cite{Kashiwaya99,ZuticValls}

\subsubsection{F/I/S-BTRS JUNCTION}
It is generally accepted that for many unconventional superconductors a subdominant component of the order parameter breaking time-reversal symmetry can be induced whenever translational symmetry is broken, e.g. near surfaces, interfaces and vortices.~\cite{Matsumoto,Sigrist98,Jacob2} For some materials, such as e.g. YBCO,~\cite{YBCO} there is controversy about the symmetry of the secondary component, namely if the order parameter is of the $d_{x^2-y^2}+is$- or $d_{x^2-y^2}+id_{xy}$-wave type. Furthermore, the splitting of the ZBCP, leading to the formation of symmetric peaks at finite bias, has been interpreted\cite{Covington97,Fogelstrom,Zhu98} as a signature of the admixture of an imaginary pair potential component with the dominant $d_{x^2-y^2}$-wave one, corresponding to a time-reversal broken symmetry state.\cite{Matsumoto,Sigrist96} The peak splitting reflects the fact that the zero-energy states are shifted by a positive or negative amount due to the Doppler shift of a finite vector potential, and the good agreement between theory and experiments suggests that the existence of BTRS is a plausible explanation for the origin of the peak splitting of the charge conductance.

\begin{figure}[!bh]
\begin{center}
\includegraphics[width=0.47\textwidth]{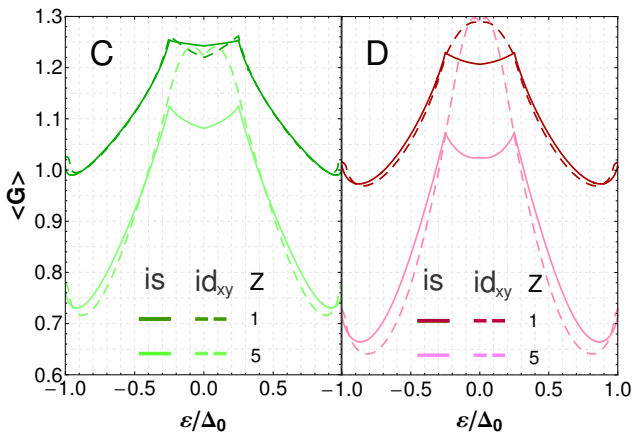}
\caption{(Color online). Averaged differential conductance
spectra for a junction with a $d_{x^2-y^2}+is$ (solid lines) and a
$d_{x^2-y^2}+id_{xy}$ (dashed lines) superconducting electrode,
evaluated at the points C (STF, left panel) and D (SBAF, right
panel) indicated in Fig.~\ref{parameterspace}, in the
intermediate ($Z$=5) and high transparency ($Z$=1) regime.
We recall that the magnetization is $M$=0.5 for both panels.} \label{BTRS}
\end{center}
\end{figure}

Thus, motivated by the fact that charge transport in junctions with a superconducting electrode could be a valuable probe of the order parameter symmetry, we compare here transport through F/I/S junctions having $d_{x^2-y^2}+is$ or $d_{x^2-y^2}+id_{xy}$ BTRS states in the S side and a SBAF or a STF in the F side. When the superconducting electrode has $d_{x^2-y^2}+is$- or
$d_{x^2-y^2}+id_{xy}$-wave symmetry, the pair potential felt by electrons (+) and holes ($-$) is $\Delta_{\sigma,\pm}^s=\Delta_1
\cos [2(\theta_\sigma'\mp\pi/4)]+i\Delta_2$ and $\Delta_{\sigma,\pm}^d=\Delta_1 \cos [2(\theta_\sigma'\mp\pi/4)]+i\Delta_2 \sin
[2(\theta_\sigma'\mp\pi/4)]$, respectively. We have analyzed spectra for several values of $\Delta_1$ and $\Delta_2$ but for brevity we show here the results only for $\Delta_1\approx0.968\Delta_0$ and $\Delta_2=0.25\Delta_0$. We notice that for this choice of $\Delta_1$ and $\Delta_2$  the gap amplitude is $\Delta_0$ for $\theta'=\pi/4$. In Fig.~\ref{BTRS} the averaged charge conductance is plotted considering the two above-mentioned BTRS superconductors for a F/I/S junction with a STF (left panel) and a SBAF (right panel), for two representative values of the barrier strength $Z$ and for a magnetization $M$ equal to 0.5. An inspection of this figure suggests that, for high $Z$, the junction exhibits for both kinds of ferromagnet a zero-bias charge response different for the two BTRS states, implying that STF/I/S or SBAF/I/S junctions are equally useful to discriminate between BTRS order parameters involved in the S-side.  For completeness, it is worth stressing that the charge conductance in the SBAF case is always larger than the one obtained in the STF one, the difference being only quantitative. In the low-barrier limit, the spectra for the two BTRS states almost coincide for a STF, while for a SBAF they are clearly more distinguishable. Therefore, we can state that in the high transparency limit a SBAF/I/S junction may be seen as a more powerful tool than a STF/I/S one to discriminate between the two BTRS states. The origin of the different behavior of the conductance at zero bias for STF and SBAF electrodes lies in an ABS at zero energy which is present in the case of $d_{x^2-y^2}+id_{xy}$-wave symmetry (only for particular angles~\cite{StefanakisABS}), but not in the $d_{x^2-y^2}+is$ one. As explained in Section II, this effect is clearly visible for a SBAF, because this kind of ferromagnetic electrode introduces an extra effective barrier which affects the charge transport of the hybrid structure, and pushing actually the junction toward the tunneling regime where ABSs become the dominant channel for transport.

\subsection{SPIN TRANSPORT}

\begin{figure}[!bh]
\begin{center}
\includegraphics[width=0.45\textwidth]{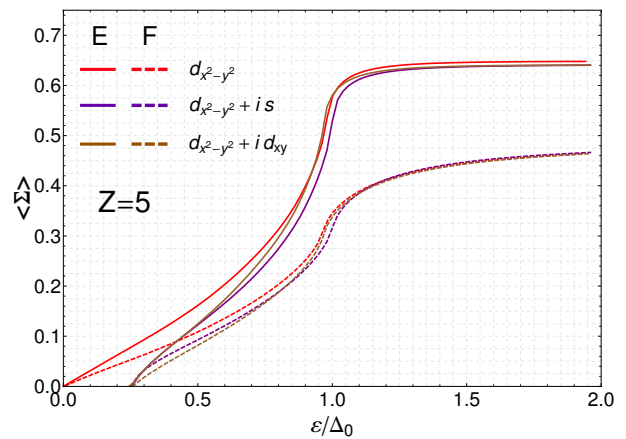}
\caption{(Color online) Averaged differential spin conductance spectra evaluated
at the points E and F reported in  Fig.~\ref{parameterspace},
for unconventional superconducting electrodes.} \label{spin}
\end{center}
\end{figure}

We have analyzed the averaged spin conductance $\langle\Sigma(\varepsilon)\rangle$ defined in Eq.(\ref{spin cond}), for the same unconventional pairing symmetries taken into account in the previous Subsection. For comparison, we have also considered a superconducting electrode characterized by a conventional $s$-wave pairing. Although several choices of the magnetization $M$ in the F electrode and of the barrier strength $Z$ have been considered, in Fig.~\ref{spin} we limit ourselves to the presentation of the spin conductance curves in the case $M=0.75$ and $Z=5$ (lower values of $Z$ and $M$ do not qualitatively alter our results). In this figure solid and dashed lines refer to the case of a junction with an STF and with a SBAF, respectively, the different colors being associated with different superconducting order parameter symmetries. For $d_{x^2-y^2}$-wave pairing, the spin conductance is non-vanishing at every finite bias and its profile exhibits, at low biases, the well-known V-shaped behavior typically produced by the gapless excitations associated with nodes of the order parameter. On the other hand, for the two BTRS states considered here the spin conductance starts being non-zero at a finite bias, corresponding to the energy of the minority component breaking time reversal, and this activated behavior is related to the nodeless properties of BTRS. Moreover, for the three kinds of unconventional pairing symmetry considered here, the spin conductance for biases lower than the energy gap $\Delta_0$ is always larger for a junction with a STF than for a junction with a SBAF. Above $\Delta_0$ this difference in magnitude gets appreciably larger, and for a given kind of ferromagnet $\mean{\Sigma(\varepsilon)}$ becomes practically indipendent on the specific pairing symmetry.

\begin{figure}[!bh]
\begin{center}
\includegraphics[width=0.45\textwidth]{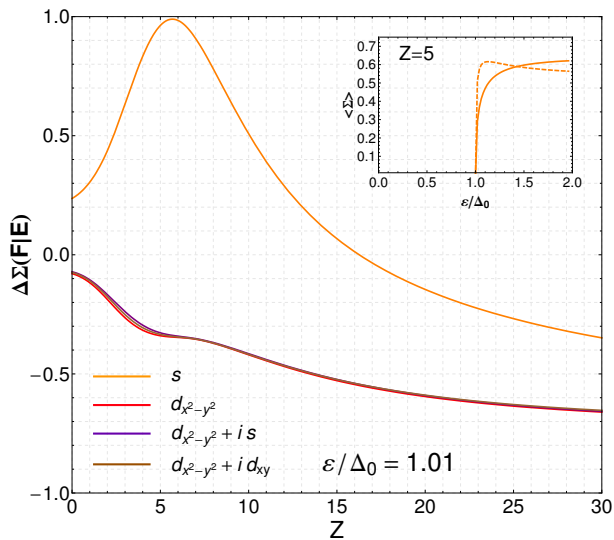}
\caption{(Color online) Relative gain in spin conductance of a
SBAF with respect to a STF, $\Delta\Sigma(F|E)=
(\mean{\Sigma(F)}-\mean{\Sigma(E)})/\mean{\Sigma(E)}$,
as a function of the barrier height $Z$, at a bias value immediately
above the energy gap $\Delta_0$, i. e. $\varepsilon=1.01 \Delta_0$.
In the inset we have plotted the spin averaged current for an
$s$-wave electrode, for the same choice of the parameters
adopted in Fig.~\ref{spin}.}
\label{delta_spin}
\end{center}
\end{figure}

Fig.~\ref{delta_spin} shows the relative gain in the spin conductance of the SBAF contribution $\mean{\Sigma(F)}$ with respect to the STF one $\mean{\Sigma(E)}$, defined as $\Delta\Sigma(F|E)=(\mean{\Sigma(F)}-\mean{\Sigma(E)})/\mean{\Sigma(E)}$, as a function of the barrier height at a fixed bias $\varepsilon/\Delta_0$=1.01 immediately above the energy gap $\Delta_0$. For comparison, we have calculated the same quantity also for the case of a junction with an $s$-wave superconductor (orange curve). We see that for a barrier height $Z$ lower than approximately 15 the gain is positive only for an $s$-wave superconductor and it can be as high as 100\%. We have checked that this peculiar effect is related to the presence of the superconducting electrode. Indeed, analyzing the spin conductance in STF/I/N and SBAF/I/N junctions, i.e. junctions where the superconductor is replaced by a normal metal, we have found that in the STF case the spin current is always greater than in the SBAF one. Looking separately at Andreev and normal reflection probabilities, we have verified that this extra spin current can be ascribed to the fact that majority electrons coming from a SBAF have a zero probability of being normally reflected at the gap edge, while electrons coming from a STF have a finite residual probability to undergo the same process. For completeness, in the inset we have reported the averaged differential spin conductance for a junction with an $s$-wave superconductor in the two cases of a SBAF (dotted line) and a STF (solid line), obtained for the same choice of parameters adopted in Fig.~\ref{spin}. We see that $\mean{\Sigma(\varepsilon)}$ is always zero below the energy gap; indeed in such situation the electrons cannot enter the superconductor side as quasiparticles because there are no quasiparticles states in the gap. Nevertheless, by Andreev reflection, they can cross the interface and decay into the Cooper pair condensate, thus preventing a spin current flow.

For spintronics applications, the ability to perform operations acting on spin currents but not on charge currents is in general highly desirable. The results presented above allow to individuate a particular situation where this is possible using F/I/S junctions with a SBAF electrode. For an $s$-wave superconductor in the case of a finite barrier strength, it has been recognized that the charge conductance is peaked around the gap edge.~\cite{Annunziata,Barsic,ZuticValls} On the other hand, we have previously shown that the spin current is zero below the energy gap $\Delta_0$ and rises abruptly just above it (inset of Fig.~\ref{delta_spin}).  If we then make the voltage across the junction vary between two limiting bias values $\varepsilon_1 < \Delta_0$ and $\varepsilon_2 > \Delta_0$ such that charge conductance is the same, it is possible to turn from a situation where a certain charge current is passing through the junction while spin current is zero ($\varepsilon = \varepsilon_1$), to a case where the spin current is different from zero and the charge conductance remains unaffected ($\varepsilon = \varepsilon_2$). Since the upper bias $\varepsilon_2$ below which the switch state is ``on" falls only slightly above $\Delta_0$, we expect that the spin current through the device will be much greater if it is generated by a SBAF rather than by a STF,  given the appreciable difference between the two cases visible in the inset of Fig.~\ref{delta_spin}.

\section{CONCLUSIONS}
In this paper we have studied the conductance spectra of ferromagnetic/insulator/superconductor hybrid structures,
developing an extension of the standard BTK approach to the case of a ferromagnetic electrode exhibiting either a standard Stoner exchange mechanism or a mass mismatch-driven ferromagnetism. We have investigated the effects induced by these two different sources of magnetization comparing the averaged charge and spin conductances of STF/I/S junctions (where only exchange splitting is present) and SBAF/I/S junctions (where only mass mismatch is present), for various symmetries of the order parameter in the superconducting electrode. Our analysis has revealed several differences between the two cases. For the charge conductance, we have found a narrower and higher peak in the SBAF/I/$d_{x^2-y^2}$ case compared to the STF/I/$d_{x^2-y^2}$ one, this finding being potentially useful for the experimental detection of a mass mismatch contribution to the magnetization.

Since the Andreev reflection is phase sensitive, the onset and amplitude of Andreev bound states, manifesting themselves in the zero bias conductance peak, is a signature of the symmetry of the order parameter. For this reason, we have also investigated the transport properties of a junction with a superconductor exhibiting a broken time-reversal symmetry of $d_{x^2-y^2}+is$ or $d_{x^2-y^2}+id_{xy}$ type. In the high transparency limit, we have found a different behavior around zero bias of SBAF/I/$d_{x^2-y^2}$+i$d_{xy}$ and STF/I/$d_{x^2-y^2}$+i$d_{xy}$ junctions, such that the use of a SBAF allows to discriminate more efficiently between BTRS states with $d_{x^2-y^2}+is$ or $d_{x^2-y^2}+id_{xy}$ pairing symmetry than STF does.
Indeed, as previously discussed a SBAF ferromagnetic electrode introduces an extra effective barrier which affects the charge transport of the hybrid structure, driving the junction toward a tunneling regime where ABSs is the dominant channel for transport.

As far as the spin transport is concerned, we have shown that the averaged spin conductance in a STF/I/S junction is greater than in a SBAF/I/S one for all the superconducting symmetries analyzed here, except for the case of a conventional $s$-wave superconducting electrode. We have also shown that a F/I/S junction with an $s$-wave superconductor can work as a switch able to turn on and off a spin current, leaving the charge current unchanged. In particular, our results show that for a wide range of interfacial barrier strengths, the spin current passing through the junction when the state of the switch is ``on" is larger if the ferromagnetic electrode is a SBAF rather than a STF. This relative increase in spin current can be very high, and for particular values of the barrier strength a gain of up to 100$\%$ can be reached.

Finally, we point out that the theoretical framework behind the calculation presented in this paper is simple enough to allow analytic solutions in the whole relevant parameter space. We leave for future work more complex approaches able to include the effect of spin-flip scattering, more realistic band structures, non-equilibrium transport, as well as a self-consistent treatment of the pair potential.

\section{ACKNOWLEDGMENTS}
We wish to thank sincerely Prof. Jacob Linder for useful discussions.


\begin{thebibliography}{99}
\bibitem{footnote} Other causes for peak splitting that do not involve magnetism are line nodes not perpendicular to the interface, i.e. $\beta\neq\pi/4$, or presence of secondary component in superconducting order parameter
breaking time reversal symmetry.
\end{thebibliography}
\end{document}